\documentclass{aa}  
\usepackage{graphicx,color}
\usepackage{txfonts}
\begin{document}
   \title{ Orbital modulation of X-ray emission lines in  Cygnus X-3.    }


  \author{O. Vilhu
         \inst{1}
          \and
          P. Hakala\inst{2}
          \and
          D.C. Hannikainen\inst{1,4}
           \and
          M. McCollough\inst{3}
           \and
          K. Koljonen\inst{3,1}
          }

  \offprints{O. Vilhu}

   \institute{Observatory,T\"ahtitorninm\"aki (PO Box 14), FI-00014 University of Helsinki, Finland\\
              \email{osmi.vilhu@astro.helsinki.fi}
         \and
             Tuorla Observatory, V\"ais\"al\"antie 20, FI-21500 Piikki\"o, Finland\\
             \email{pahakala@astro.utu.fi}
             \and
             Harvard-Smithsonian Center for Astrophysics, 60 Garden Street, Cambridge, MA 02138, USA\\
             \email{mccolml@head-cfa.harvard.edu}
             \and
             Mets\"ahovi Radio Observatory/TKK, Mets\"ahovintie 114, FI-02540 Kylm\"al\"a, Finland\\
             \email{diana@kurp.hut.fi}
                 }

   \date{Received 05/11/2008; accepted 19/04/2009  }

 
  \abstract
  {}
   {We address the problem where the X-ray emission lines are formed and  investigate orbital dynamics   using $\it{Chandra}$ HETG observations, photoionizing calculations  and numerical wind-particle simulations.
The aims were to set constraints on the masses of the components of this close binary system consisting of a Wolf-Rayet (WR) star and a compact component and to investigate the nature of the latter (neutron star or black hole). The goal was also to investigate P Cygni signatures in line profiles.
}
   {The observed \ion{Si}{xiv} (6.185 \AA) and \ion{S}{xvi} (4.733 \AA) line profiles at four orbital phases were fitted with P Cygni-type profiles consisting of an emission and a blue-shifted absorption component.  Numerical models were constructed using photoionizing calculations  and  particle simulations. In the models, the emission originates in the photoionized wind of the WR companion illuminated by  a hybrid source: the X-ray radiation of the compact star   and the photospheric EUV-radiation from the WR star. 
   }
   {Spectral lines with  moderate excitation (such as \ion{Si}{xiv} and \ion{S}{xvi})  arise in the  photoionized wind. The emission component exhibits   maximum blue-shift at phase 0.5 (when the compact star is in front), while the  velocity of the absorption component is constant (around $-900$ km/s). Both components, like the continuum flux,  have intensity maxima around phase 0.5.    The simulated \ion{Fe}{xxvi} Ly$\alpha$ line (1.78 \AA, H-like)  from the wind is weak compared to the observed one. We suggest that it originates in the vicinity of the compact star, with a maximum blue shift at phase 0.25 (compact star approaching).    By combining the mass function derived with that from the infrared \ion{He}{i} absorption (arising from the WR companion), we constrain the masses and the inclination  of the system.
 } 
   {The \ion{Si}{xiv} and \ion{S}{xvi} lines and their radial velocity curves can be understood in the framework of a photoionized wind involving a hybrid ionizer. Constraints on the compact star mass and orbital inclination ($i$) are given using the mass functions derived from the \ion{Fe}{xxvi} line and \ion{He}{i} 2.06 $\mu$m absorption. Both a neutron star at large inclination ($i\ge{60}$ degrees) and a black hole at small inclination
are possible solutions. The radial velocity amplitude of the \ion{He}{ii} 2.09 $\mu$m emission (formed in the X-ray shadow behind the WR star) suggests $i=30$ degrees, implying a possible compact star mass between 2.8 -- 8.0 M$_{\sun}$. For $i=60$ degrees the same range is 1.0 -- 3.2 M$_{\sun}$. }


   \maketitle
%

\section{Introduction}

Cygnus X-3 (4U 2030+40, V1521 Cyg) is a high-mass X-ray binary (HMXB) located
at a distance of 9 kpc with a close binary orbit (P = 4.8 hour; Liu \cite{HMXB}, Hanson \cite{hanson}). 
The compact star is either a neutron star or a black hole and the companion is a WN5--7 type Wolf-Rayet (WR) star (van Kerkwijk \cite{kerk92},\cite{kerk96}),    {with strong evidence pointing to an early-type WN star (Fender et al. 1999).
The system appears to be engulfed in an extended envelope produced by the wind of the WR star.
This notion is supported by the fact that 
massive winds are generally observed in Wolf-Rayet stars (Langer \cite{langer}, Crowther \cite{crowther}).} 

   {Specifically with regard to Cyg X-3, large and complex absorption is found in broadband X-ray spectral studies (e.g. Vilhu \cite{vil}, Hjalmarsdotter \cite{nea}, Szostek and Zdziarski \cite{anna}), the orbital modulation of  X-ray light curves can be attributed to electron scattering absorption in an asymmetric wind (with respect to the X-ray emitter, Pringle \cite{pringle}, Willingale \cite{willingale}), and the source exhibits a very rich $\it{Chandra}$  HETG emission line spectrum (Paerels \cite{paerels}).
The spectrum shows radiative recombination continuums (RRC) that is a clear indication of a photoionized plasma (an alternative to a collisionally ionized gas). 
Liedahl et al. (\cite{lied}) and McCollough et al. Ê(\cite{mike}) report P Cygni profiles in some X-ray emission lines. 
Stark and Saia (\cite{stark}) found Doppler modulation of X-ray lines in
$\it{Chandra}$ data from April 2000 when Cyg X-3 was in the high state,  
and used this to constrain the compact star mass to below 3.6 solar masses. }

In the present paper we analyse one long $\it{Chandra}$ HETG observation
taken during a high state of Cyg X-3. 
We fit the \ion{Si}{xiv} 6.185 $\AA$  Ly$\alpha$-line at four orbital phases with a P Cygni-type 
profile (an emission component with a blue-shifted absorption component). 
Results for the \ion{S}{xvi} Ly$\alpha$-line at 4.73 $\AA$, the FeÊ    {Ly$\alpha$ line at 1.78 $\AA$, and the He$\alpha$ line at 1.859 $\AA$ are also presented}. 
Using  a two-component hybrid ionizing flux arising from the compact object and
the WR star
and  the XSTAR photoionizing code (Kallman \cite{xstar}; for a review of photoionized winds see also Liedahl \cite{liedahl}), we  estimate line emissivities in the radiation field. 
Radial velocities and lightcurves were computed  from a wind model based on particle simulations in the binary potential and compared with those we observed.
We conclude by discussing
constraints on the orbital inclination, component masses, and wind velocity.        
 
\section{   Chandra observations}

   {Chandra High Energy Transmission Grating (HETG) observations (OBSID: 7268, PI:
M. McCollough) were scheduled on 2006 January 25--26 as a part of a
large international campaign to study Cyg X-3 during an active state.
The observations started on MJD 53760.59458 (when the source
was at phase 0.053) and went through to 
MJD 53761.42972 (phase 4.240).}
During these observations Cyg X-3 was in a high soft state (quenched radio state)
with an average $\it{RXTE}$/ASM count rate around 30 cps (corresponding to 400 mCrab;
typical hard state count rates are less than 10 cps). 

The HETG consists of two gratings,  
the Medium Energy Grating (MEG) and the High Energy Grating (HEG).
The observations were done in CC mode (continuous clocking) with a window
filter applied to the zeroth order.  
The CC mode was used to avoid pileup in the
dispersed spectra and the window filter was applied to avoid telemetry
problems that could result from excessive counts in the zeroth order.  
A measure of pileup is the count rate per frame time (time to read out the chip) in a 
three-by-three pixel island.  
For this data set this quantity was always less than 0.02, and
from Poisson statistics this implies a pileup rate of less than 1 per cent.  
The downside of using this mode is that spatial information is lost and the background is 
an order of magnitude higher.

During spectral extraction regions, either side of the dispersed spectra were
used to create a background spectrum to be used in the analysis.  
In addition to instrumental background components this background was also used
in the removal of components due to scattering.

%

The observation lasted over four orbital cycles and 
was divided into 4 and 10 phase groups. 
For each group, HEG and MEG spectra with their responses were extracted and first-order spectra used in the present study. The spectral resolutions were 0.023 $\AA$ and 0.012  $\AA$ for MEG and HEG, respectively (corresponding to 1160 km/s and 580 km/s at the \ion{Si}{xiv} line).

The spectra  were analysed with  XSPEC (version 11) and extracted into a format which could be used by programs written for IDL (version 6.2).
We concentrated on the \ion{Si}{xiv} 6.185 $\AA$  Ly$_{\alpha}$-line as
it is free of blends and the line profiles should be reliable. 
However, profile fits for the \ion{S}{xvi} Ly$\alpha$ line at 4.73 $\AA$ are also included in this work 
as well as the radial velocity study of the iron  lines at 1.78 $\AA$ and 1.86 \AA.

\section{XSTAR simulations}

XSTAR is a computer code that calculates radiation effects in a cloud surrounding an ionizing source (Kallman \cite{xstar}); in particular, the  complex emission line spectrum of different species is computed. 
The important input parameters are the chemical composition, density, and radius of the spherical cloud, 
and the X-ray/EUV spectrum of the central source. 
Of particular relevance is the radiation (ionizing luminosity) above the 
hydrogen ionization limit of 13.6 eV.
The code was run in constant density mode, and the temperature was calculated
to be the equilibrium value.
The input parameters are listed in Table~\ref{XSTARMODEL} and explained below.                         

\begin{table}
\begin{minipage}[t]{\columnwidth}
\caption{The ionizing sources and gas parameters used in the XSTAR code.} 
\label{XSTARMODEL}
\centering
\renewcommand{\footnoterule}{}  
\begin{tabular}{lccccc}
\hline 
 & & \\
compact star& WR star  &  gas envelope    \\
 & & \\
\hline
 & & \\
 compPS&BB &N$_H$=5$\times$ 10$^{23}$ cm$^{-2}$ \\
 T$_{bb}$=0.71 keV&T$_{eff}$ = 0.01 keV  & N=10$^{11}$ cm$^{-3}$   \\
  L=2.46$\times$10$^{38}$ erg/s & L=5$\times$ 10$^{38}$ erg/s& H=0.1,He=10,CNO\\
\hline
\end{tabular}
\end{minipage}
\end{table}

Since the source of the wind
is a WN(4--7) star (van Kerkwijk et al.  \cite{kerk96}) we used helium-rich gas abundances 
with CNO equilibrium values: H=0.1, He=10, C=0.56, N=40, O=0.27; the other 
elements were set to solar, i.e. 1 (Crowther \cite{crowther},Vilhu \cite{moscow}). 
The role of abundances is not significant, however, since we are not concerned
with absolute values of fluxes, abundances or densities.
The column (N$_H$ = 5$\times$10$^{23}$ cm$^{-2}$) and particle density (N=10$^{11}$ cm$^{-3}$) were chosen resulting in a cloud of size $\sim 1$ AU.
The N$_H$-value is compatible with the Thomson scattering origin of the orbital modulation of the 
hard X-ray light curves (Vilhu et al. \cite{moscow}).  

Both binary components were used as the ionizing radiators, thus forming a hybrid ionizer. 
To obtain initial estimates for the parameters of the X-ray ionizing source (i.e. the compact star), we fit {\it RXTE}/PCA spectra (xp5100, xp5200, xp5300) that were simultaneous to the {\it Chandra}
observation in XSPEC with a Comptonizing model (compPS, Poutanen \& Svensson 1996) attenuated by 
photoelectric absorption (wabs).
For a distance of 9~kpc, the resulting unabsorbed bolometric luminosity is 
2.46$\times$10$^{38}$ erg/s.
The model includes photoelectric absorption (8$\times$10$^{22}$ cm$^{-2}$), a black body radiator (T$_{bb}$ = 0.71 keV), a Comptonized component (with kT$_e$=5.8 keV and $\tau$=2.0) and a reflection component (with relative reflection = 4.5).  
The spectrum is soft and 
dominated by a disc black body, typical for X-ray binaries in the high state. 
Note that we do not claim that this model is the best one to represent Cyg X-3 in the high state -- the true absorption is probably more complex than is assumed here and the compPS model too simple. 
   {However, the model does give a viable
numerical framework for our present purposes, 
including a correct value for the ionizing luminosity.}

\begin{figure}
   \centering
   \includegraphics[width=9cm]{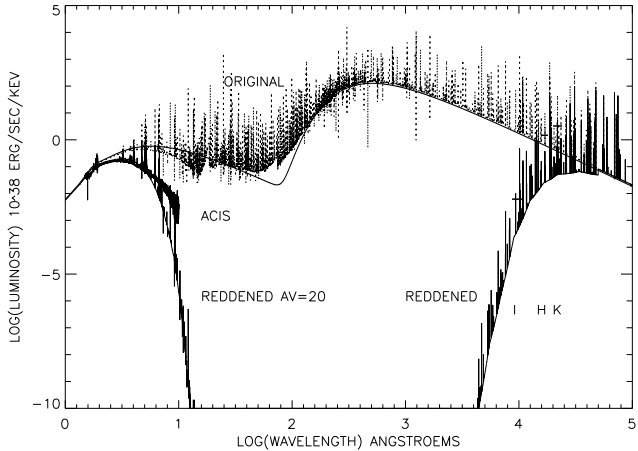}
      \caption{The outgoing broadband spectrum. 
      The dotted line traces the original spectrum and the solid line the reddened spectrum 
      as computed with the hybrid ionizer
      (see the text and Table~\ref{XSTARMODEL}). 
      The observed IR magnitudes (Hanson et al. \cite{hanson}, short horizontal lines) and the 
      mean {\it Chandra} HETG spectrum (OBSID 7268) are shown. 
      The solid continuous line is the spectrum of the incident hybrid source.}
         \label{spsbroad}
   \end{figure}

\begin{figure}
   \centering
   \includegraphics[width=9cm]{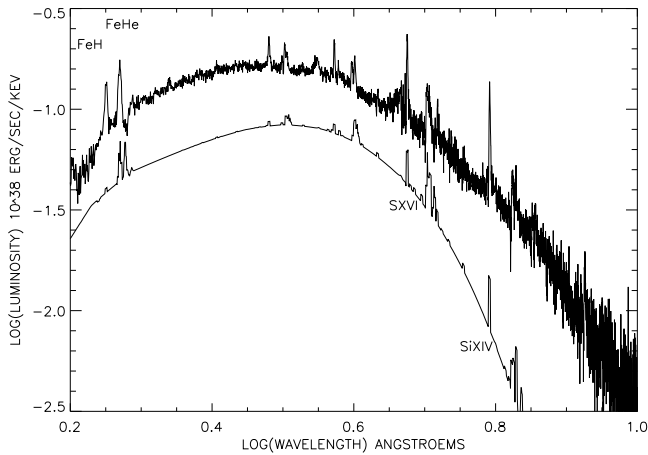}
      \caption{A zoom of the simulation of Fig.\ref{spsbroad} (shifted vertically by $-$0.4) overplotted with the mean $\it{Chandra}$ HETG-spectrum (OBSID 7268).  }
         \label{spszoom}
   \end{figure}

For the WR star we used values similar to those of V444 Cygni, which is a WN5 + O6 eclipsing binary:  a mass of 10 M$_{\sun}$, a radius of 0.92 R$_{\sun}$, a luminosity of 10$^{5.1}$ L$_{\sun}$ and T$_{eff}$ = 115 000 K (Langer \cite{langer}). 
We approximate the spectrum with a black body of T$_{eff}$ = 0.01 keV (116 000 K) and bolometric luminosity of 5$\times$10$^{38}$ erg/s; the ionizing luminosity is then 4.47$\times$10$^{38}$ erg/s. 
The incident and outgoing fluxes are illustrated in Figures ~\ref{spsbroad} and ~\ref{spszoom}. 
The outgoing flux was reddened by A$_V$ = 20 and A$_V$/N$_H$ = 5.3$\times$10$^{-22}$ cm$^{2}$, 
following Tables 21.6 and 21.7 in Cox (\cite{allen}).

   {Figures ~\ref{spsbroad} and \ref{spszoom} show that even though the model outlined above 
adequately fits the observed $\it{Chandra}$ HEG spectrum, it does not fully reproduce the 
infrared (IR) continuum and there is a soft X-ray excess visible relative to the continuum.
Depending on whether Cyg X-3 is in quiescence or in outburst, the infrared magnitudes can
vary by a magnitude (Hanson et al.\ 2000).
Fig.\ref{spsbroad} shows the infrared magnitudes given in Hanson et al. (\cite{hanson}): m$_H$=13.1$\pm{0.1}$ 
and m$_K$=11.7$\pm{0.1}$, based on observations during quiescence by 
Fender et al. (\cite{fender}).  
The soft X-ray excess is likely due to background issues and/or an inadequate
representation of the  absorption in the $\it{RXTE}$/PCA spectra. 
However, these discrepancies do not affect the results presented here.}

\begin{figure}
   \centering
   \includegraphics[width=15cm]{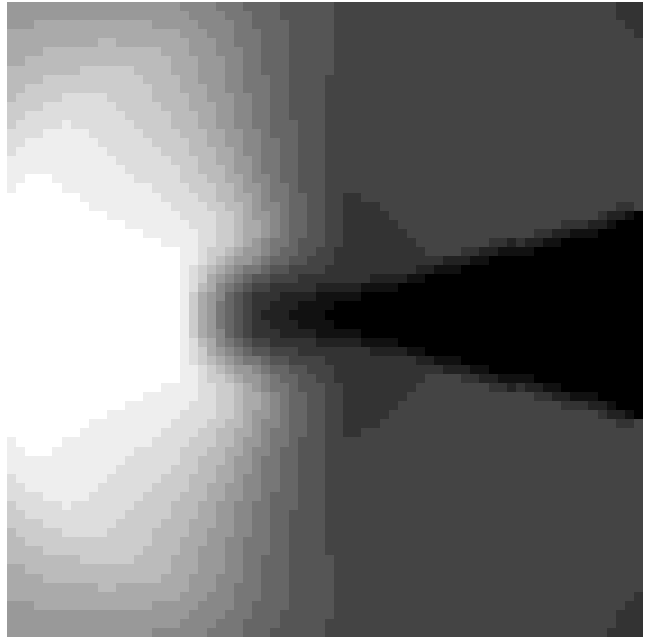}
      \caption{\ion{Si}{xiv} Ly$\alpha$-line emissivity per unit volume for  constant density N=10$^{11}$ cm$^{-2}$ (on orbital plane) based on the hybrid ionizer (Table ~\ref{XSTARMODEL}). The plot size is 32 R$_{\sun}$, which corresponds to ten binary separations. The X-ray shadow of the WR-companion is seen on the right. The compact component is in the dark spot slightly to the left of figure center. In this grey scale  white corresponds to  maximum and black to zero emissivity. 
      At orbital phases 0.0 and 0.5 the system is viewed from the right and left, respectively.     }
         \label{emissi}
   \end{figure}

XSTAR requires as input a single ionizing source in the cloud center, 
although in our case we have two distinct ionizers. 
Hence, several runs with different He-star luminosities and distances from the compact object  were 
performed (using spherical shells) so as to interpolate the correct radiation field at any spatial point where 
the line emissivity was computed (see Fig.~\ref{emissi} for \ion{Si}{xiv}).

\begin{figure}
   \centering
   \includegraphics[width=15cm]{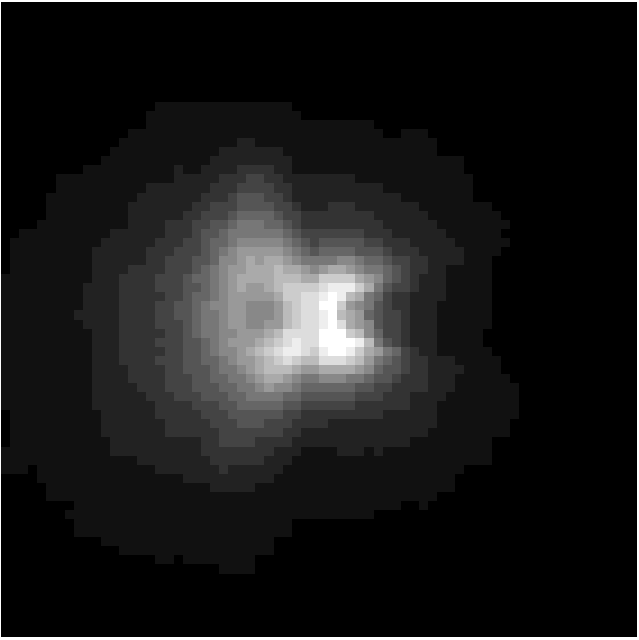}
      \caption{\ion{Si}{xiv} Ly$\alpha$ line emission in the photoionized wind model with q=5 and V=2000 km/s on the orbital plane. The plot size is 32 R$_{\sun}$ (ten binary separations). The grey scale is linear with white corresponding to maximum and black to zero emission. At orbital phases 0.0 and 0.5 the system is viewed from the right and left, respectively. }
         \label{pvolemis}
   \end{figure}

Additional emissivity grids were computed by varying the particle density. 
The degree of photoionization is governed by the density of the gas through the 
the ionization parameter, which is proportional to the ionizing luminosity and inversely proportional to distance squared and particle density, $\xi$=L/(nR$^2$).
Ionization is sensitive to the density of the gas (which itself depends on the ionizing species in 
question): it can increase or decrease in accordance.
The final line emission was then computed in the wind model by fixing the mean density 
to 10$^{11}$ cm$^{-3}$ (see Fig.~\ref{pvolemis} for \ion{Si}{xiv}). 
This mean density gives the correct electron scattering opacity along the line of sight and produces a partial eclipse 
in the light curve by a factor of two, as observed.  
A similar procedure was undertaken
to estimate the absorption in the transmitted spectrum.
 
The emission scales relative to the equilibrium temperature as T$^{-0.6}$,
 which explains 
the enhanced emission when the extra heater, the WR star, is farther away (see Fig.~\ref{emissi}).
The line absorption in the transmitted flux, in turn, has no temperature dependence. 

%

%
%

\section{ Particle simulations}

For  modeling the velocity and density fields so as to obtain a reasonable input model for XSTAR, test particles were injected symmetrically from the Wolf-Rayet companion and their trajectories were followed in the binary potential. 
Neither acceleration nor interaction between the particles was assumed,
which is a good assumption for the supersonic velocity of the wind.
Typical values  for the wind velocity and mass loss of WR stars are 2000 km/s 
(observed wind velocities lie in the range 1000--3000 km/s (Cox \cite{allen}))
and  3$\times$10$^{-5}$ M$_{\sun}$/year, 
respectively (Langer \cite{langer}). 
The masses of WN stars are between 10 -- 80 M$_{\sun}$, while the range of WC stars is narrower 
(9 -- 16 M$_{\sun}$, Crowther \cite{crowther}). 

In our simulations we used  wind velocities between 1200 -- 2200 km/s, a mass of 10 
M$_{\sun}$  for the WR star (with radius 0.92R$_{\sun}$, Langer \cite{langer}), 
and 2 M$_{\sun}$ and 5 M$_{\sun}$ for the compact star (see Table~\ref{syst}). 
We used 16-s time steps and the number of particles within the escape shell, 30$\times$separation, 
was kept constant at 10$^6$. 
A new wind particle was always generated when a test particle moved outside 
the escape shell, returned back to the WR star, or passed closer than 0.15 R$_{\sun}$ from the 
compact star (the assumed accretion radius). 
In all cases a stable  structure was formed after about five binary revolutions. 

%
\begin{table}
\begin{minipage}[t]{\columnwidth}
\caption{Binary parameters used for wind particle simulations with
wind velocities V = 1200 -- 2200 km/s.} 
\label{syst}
\centering
\renewcommand{\footnoterule}{ }  
\begin{tabular}{lccccc}
\hline 
 & & \\
M$_{WR}$ & M$_C$ & mass ratio q    & separation a & V$_C$ & V$_{WR}$      \\
 (M$_{\sun}$)& (M$_{\sun}$)& (M$_{WR}$/M$_C$)& (R$_{\sun}$) & (km/s) &(km/s) \\
\hline
 & & & & & \\
 10 & 2 & 5 & 3.2 & 691.0 &138.2  \\
 10 & 5 & 2 & 3.4 & 595.5&297.7  \\
\hline
\end{tabular}
\end{minipage}
\end{table}

At large distances (larger than the binary orbit) the simulated 
outward directional velocity and particle density  
as a function of the distance, r, from the WR component,
follow quite closely the semiempirical laws:  
\\

V = const(1-R$_{wr}$/r)$^{\beta}$ ; N = const/[r$^2$$\times$(1-R$_{wr}$/r)$^{\beta}$]
; $\beta$=2.

\section{Line profile fitting results}

\subsection{P Cygni model}

As already remarked in McCollough et al. (\cite{mike}) and Liedahl et al. (\cite{lied}) the first HETG  
spectra of Cyg X-3 showed P Cygni-type line profiles (see also Sako et al. \cite{velax1}). 
This is evident also from our phase-resolved \ion{Si}{xiv} spectra. 
We modelled the line profile with two gaussians, one representing emission (gauss1) 
and the other absorption (gauss2), 
as follows:
\\

F = exp($-$gauss2)$\times$(gauss1 + continuum),  where 
\\

 gauss = A0$\times$exp{$-$0.5$\times$[($\lambda$-A1)/A2]$^2$} .
\\

The A-parameters are constants obtained in the fitting procedure and $\lambda$ is the wavelength. 
The continuum was kept constant over the line profile. 
In practice, $\lambda$ was replaced by the velocity
($\delta$$\lambda$/$\lambda$$\times$c)
measured from the line center. This approach is an approximation
of the P Cygni profile. The shape of the feature is a function of the distribution of wind velocities and is  more complicated than the simple model above.

The observed profiles were fit using the model described above with  the IDL procedure mpcurvefit.pro (written by Craig Markwardt). 
Normal weighting was used (weights = 1/error$^2$, where `error' is the observation 
error of the flux inside a wavelength bin). 
The line profile was treated within $\pm$ 4000 km/s from the line center, 
corresponding to 66 HEG  and 33 MEG $\lambda$-bins at the \ion{Si}{xiv} line, respectively. 
In the final fit  the absorption line central velocity was frozen to  $-900$ km/s, to avoid artificial 
jumps in some parameters. 
However, this value is close to what is found when all seven parameters were left free. Mean intrinsic FWHM's of the absorption and emission profiles were 750$\pm{200}$ km/s and 1850$\pm{200}$ km/s, respectively. 
Due to the requirement of high S/N, we were restricted to using only four phase bins.

\subsection{Radial velocities}

\subsubsection{The  \ion{Si}{xiv} line}

\begin{figure}
   \centering
   \includegraphics[width=9cm]{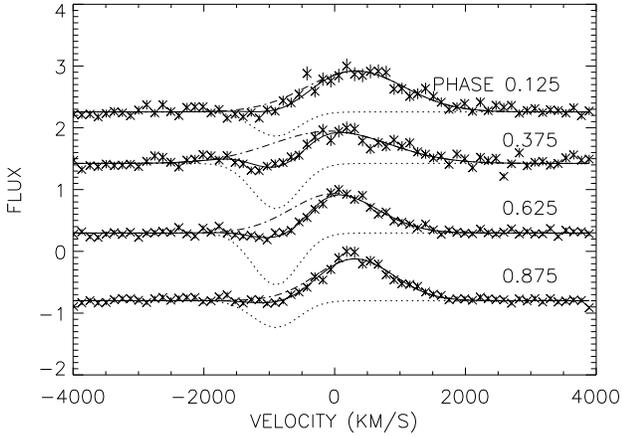}
      \caption{\ion{Si}{xiv} 6.1850 $\AA$  Ly$\alpha$ line profiles at four orbital phases (HEG 1st order)  overplotted with  P Cygni-fits. The profiles are scaled with the  maximum value and shifted vertically by a constant amount. The solid line gives the fit, dash-dot line  the emission component and the dotted line shows the absorption. The x-axis velocity is measured from the line centrum. }
         \label{profilesnormheg}
   \end{figure}

\begin{figure}
   \centering
   \includegraphics[width=9cm]{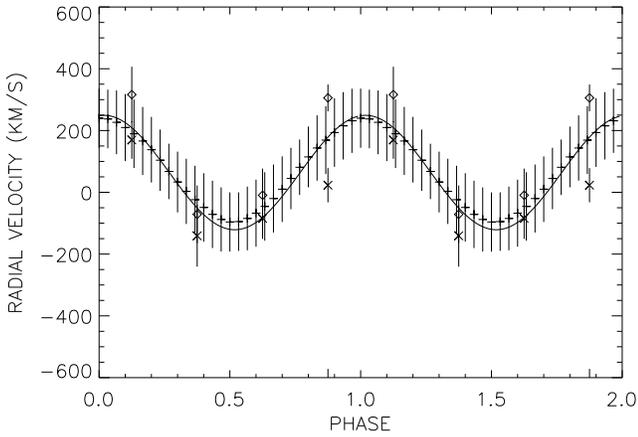}
      \caption{   {Radial velocity curve of the \ion{Si}{xiv} Ly$\alpha$ emission component  at four orbital phases overplotted with the model q=5 and V=2000 km/s for inclination 60 degrees (dashed line with error bars).
     The data are represented by symbols with error bars: MEG=crosses, HEG=diamonds.   }  }
         \label{acis7268radvel}
   \end{figure}

The  \ion{Si}{xiv} Ly$\alpha$ line profile fitting results are shown in Fig.~\ref{profilesnormheg} 
and Table~\ref{radvel}. 
   {In Fig. 5 the dash-dot line shows the emission component (continuum + gauss1) 
while the dotted line shows the absorption (continuum $-$ gauss2)}. 
In Table~\ref{radvel} we also include the \ion{S}{xvi} Ly$\alpha$-line. 
The $\chi^2$ values from the \ion{S}{xvi} line fits are somewhat smaller 
than those of the \ion{Si}{xiv} line -- this is due to larger observational errors in the latter. 
Fig. ~\ref{acis7268radvel} compares the observed \ion{Si}{xiv} emission radial velocity curve   with the model (q=5, V=2000 km/s, i = 60 deg). 
The rather conservative error bars were computed as errors in the sine curve fit to the 
mean MEG+HEG data (the solid curve).

\begin{table}
\begin{minipage}[t]{\columnwidth}
\caption{ Radial velocity parameters for the  emission component of the \ion{Si}{xiv} (6.185 \AA) and \ion{S}{xvi} (4.733 \AA) Ly$\alpha$ lines and mean $\chi^2$-values of P Cygni profile fits. }
\label{radvel}
\centering
\renewcommand{\footnoterule}{}  
\begin{tabular}{lccccc}
\hline 
 & & \\
line&$\Gamma$ & K  & bluest& MEG  & HEG   \\
 & & & & \\
  &km/s&km/s & phase & $\chi^2$(DOF) & $\chi^2$(DOF) \\
\hline
 & & & & \\
 \ion{Si}{xiv}&64$\pm{38}$&185$\pm{52}$&0.52$\pm{0.04}$&94(26)&62(59) \\
 \ion{S}{xvi} &75$\pm{33}$&100$\pm{36}$&0.44$\pm{0.08}$&34(26)&40(59) \\
 
 & & & & & \\
\hline
\end{tabular}
\end{minipage}
\end{table}

The observed radial velocity semiamplitude of the \ion{Si}{xiv}  emission component 
was 185 $\pm{52}$ km/s with a Gamma-velocity of 64$\pm{38}$ km/s 
and maximum blue-shift occurring at phase 0.5.  
The \ion{S}{xvi} Ly$\alpha$ line at 4.73 $\AA$ behaves in a rather similar way but
with a somewhat smaller radial velocity amplitude (see Table~\ref{radvel}).
 
The maximum blueshift at phase 0.5 is due to the enhanced emissivity behind the X-ray source 
(Figs.~\ref{emissi} and ~\ref{pvolemis}). 
To explain the radial velocity amplitude observed, high wind velocity is required.
The effects of inclination, wind velocity and mass ratio are compared in Table~\ref{radvelcomp}, 
where $\chi^2$-values  between the observed and model radial velocity curves  are computed 
for different input parameter values. 
Due to the  conservative errors used,  the $\chi^2$-values are relatively small. 
If the mean errors of the observations were to be used, then the $\chi^2$ values would be 
larger by a factor of 5.2.

\begin{table}
\begin{minipage}[t]{\columnwidth}
\caption{ Effects of inclination, wind velocity and mass ratio  on the observed and computed \ion{Si}{xiv} radial velocity curves.  }
\label{radvelcomp}
\centering
\renewcommand{\footnoterule}{}  
\begin{tabular}{lccccc}
 & & & & \\
 inclination &V$_{wind}$& q & $\chi^2$  \\
\\
degrees &km/s & &DOF=28  \\
\\
\hline
\\
 60&2000 &5 &0.84  \\
45&2000 &5 &3.7  \\
 30&2000 &5 &13.5  \\
 60&1800 &5 & 19.9 \\
 60&2000 &2 &6.73  \\
 & & &  \\
\hline
\end{tabular}
\end{minipage}
\end{table}

\begin{table}
\begin{minipage}[t]{\columnwidth}
\caption{Observed (OBSID 7268 HEG) and  model (q=5, V=2000 km/s) line luminosities (in units of 10$^{35}$ erg/s).       }
\label{lineint}
\centering
\renewcommand{\footnoterule}{}  
\begin{tabular}{lccccc}
\hline 
 & & \\
ion&wavelength (\AA) &Observed Lum & Model Lum      \\
 & & & \\
\hline
 & & \\
 \ion{Si}{xiv}&6.185&1.00$\pm{0.08}$ & 1.30 \\
 \ion{S}{xvi}&4.733&1.80$\pm{0.10}$& 1.60 \\
 \ion{Fe}{xxv} &1.859& 7.50$\pm{0.60}$ &2.10 \\
 \ion{Fe}{xxvi} &1.780& 6.05$\pm{0.50}$ &0.90 \\
 & & & & & \\
\hline
\end{tabular}
\end{minipage}
\end{table}

\subsubsection{The iron lines}

In the Fe-line complex at 6.6 -- 6.9 KeV, the dominating H- and He-type lines are 
resolved
in the HETG-spectrum (see Fig.~\ref{spszoom}). 
The lines are formed close to the compact star in a region of high ionizing flux 
and require special attention. 
In particular, our ionizing wind models predict a much too weak \ion{Fe}{xxvi}  line, 
as can be seen in the large discrepancy between the observed line luminosity   
and the model (see Table ~\ref{lineint}). 
In order to account for this discrepancy, an extra radiation source is required. 
The He-type \ion{Fe}{xxv} line is a composite of  forbidden (1.869 \AA), intercombination (1.859 \AA) and resonance (1.852 \AA) lines arising partially in the wind (with intensity ratios 0.35:0.45:0.19) and partially in the region where the \ion{Fe}{xxvi} line originates.

\begin{figure}
   \centering
   \includegraphics[width=9cm]{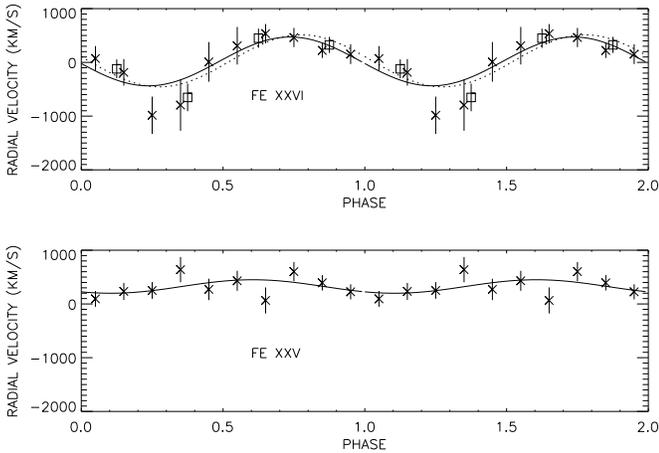}
      \caption{   {Upper plot: Radial velocity curve of the \ion{Fe}{xxvi} (H-like) Ly$\alpha$-line   overplotted with the sine curve fit using 10 phase bins (crosses and the solid line). For comparison the same is shown when the observation is divided into 4 phase bins (sqaures and the dotted line).  Lower plot: The 10 phase bin case for \ion{Fe}{xxv} (He-like). Lines were treated as single gaussians. Phases 0.0 and 0.5 correspond to the cases when the WR star and compact star, respectively, are in front.} }
         \label{FeHradvel}
   \end{figure}

To study  the radial velocity curves we divided the observation into four and ten phase bins. 
The line centroids and fluxes were determined by  single Gaussian fitting in the HEG spectrum. 
No absorption component was detected, since the spectral resolution is over three times worse than in the \ion{Si}{xiv} line region.  
The wind model predicts for the Fe lines similar radial velocity 
curves to the cooler \ion{Si}{xiv} and \ion{S}{xvi} ions, but in fact the observed 
radial velocities don't correspond to the model.
For \ion{Fe}{xxvi}, the radial velocity analysis was performed
separately for the four and the ten phase bins. 
In addition, we eliminated
the two most blueshifted phase bins and undertook a third  study without them (the eight bin case). 
The results are shown in Fig.~\ref{FeHradvel} and Table~\ref{FeHradialvel}.

\begin{table}
\begin{minipage}[t]{\columnwidth}
\caption{ Parameters for the radial velocity curves of  \ion{Fe}{xxvi} (1.78 \AA) and \ion{Fe}{xxv} (1.86 \AA) lines (treated as single gaussians and using different numbers of phase bins).}  
\label{FeHradialvel}
\centering
\renewcommand{\footnoterule}{}  
\begin{tabular}{lccccc}
\hline 
 & & & & & \\
line & bins &$\Gamma$ & K  & bluest& red$\chi^2$(DOF)  \\
 & & & & & \\
  & &km/s&km/s & phase &    \\
\hline
 & & & & & \\
 \ion{Fe}{xxvi}&4 &28$\pm{95}$&484$\pm{133}$&0.27$\pm{0.04}$& 2.85(1) \\
 \ion{Fe}{xxvi}&10 &17$\pm{86}$&454$\pm{108}$&0.23$\pm{0.04}$& 0.96(7) \\
 \ion{Fe}{xxvi}&8 &102$\pm{92}$&317$\pm{126}$&0.24$\pm{0.04}$& 0.51(5) \\
 \ion{Fe}{xxv}&10&324$\pm{56}$&125$\pm{78}$&0.10$\pm{0.09}$& 0.98(7) \\
 & & & & & \\
\hline
\end{tabular}
\end{minipage}
\end{table}

The maximum blueshift of the \ion{Fe}{xxvi} line  occurs around phase 0.25, when
the compact star is approaching.
Hence, it is tempting to associate the line-forming region with the compact component, 
presumably its accretion disc or corona. 
If the fit for \ion{Fe}{xxvi} is forced to result in a maximum blueshift at exactly phase 0.25, the resulting K-amplitudes do not deviate significantly from those given in Table~\ref{FeHradialvel}.

Despite the apparent complexity in producing the \ion{Fe}{xxvi} line,
we assume that it originates in the vicinity of the compact object, primarily because the
maximum blueshift occurs at phase 0.25 when the compact object is approaching.
 
For \ion{Fe}{xxvi} we compromise and adopt 
the mean K value in Table~\ref{FeHradialvel} (K = 418 $\pm{123}$ km/s) 
to compute the mass function using the definition
 \\
  
 f = K$^3$P/(2$\pi$G) = M$^3$$\times$sin(i)$^3$/M$_{tot}^2$
 \\
 
\noindent for a circular orbit. 
From this we get  f$_{WR}$ = 1.51$^{+1.75}_{-0.99}$ M$_{\sun}$.
All  three 
cases for \ion{Fe}{xxvi} lie, however, within the error ranges.
To check this, we also analysed another high state observation of Cyg X-3 (OBSID 6601) 
that we divided into ten phase bins. 
The result for \ion{Fe}{xxvi} was consistent with
those of Table~\ref{FeHradialvel}: K = 470 $\pm{130}$ km/s when the maximum blueshift 
was frozen to phase 0.25 (reduced $\chi^2$ = 1.3).

The He-like \ion{Fe}{xxv} line is more problematic. 
It is formed  partially in the photoionized wind, like \ion{Si}{xiv}, and in the vicinity of the compact 
object, like \ion{Fe}{xxvi}. 
However, extra complications  arise due to its triplet nature 
(resonance, intercombination and forbidden components) which may be selectively 
absorbed as a function of the orbital phase, and hence we do not treat it 
further here.


%

\subsection{Light curves of the emission and absorption components}

\begin{figure}
   \centering
   \includegraphics[width=9cm]{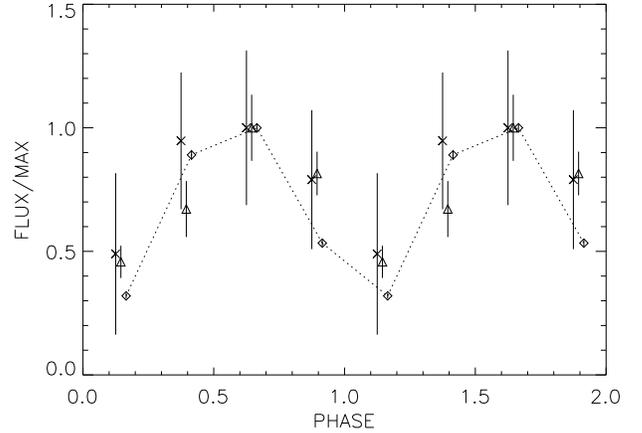}
      \caption{  {The integrated area of the \ion{Si}{xiv}-line  absorption component   is represented by crosses, the flux of the emission component  by triangles.  The diamonds joined by the dotted line gives the continuum at the \ion{Si}{xiv} line. All parameters are scaled by their maximum values.}   }
         \label{taus}
   \end{figure}

   {
In Fig.~\ref{taus} the integrated area of the \ion{Si}{xiv} line absorption component (integrated gauss2) is represented by crosses (from the mean MEG+HEG data) scaled by its 
maximum value (525 km/s). 
The continuum (scaled by 5.0 $\times$ 10$^{36}$ erg/s/keV) and the emission component 
(scaled by 1.5 $\times$ 10$^{35}$ erg/s) are represented by diamonds and triangles, respectively. }

   {Approximately one half of the \ion{Si}{xiv} line  emission originates from the 
far
side of the X-ray source (compact star), 
accounting for
the roughly similar orbital modulation of the line photons and 
the continuum. 
The wind particles themselves are relatively
symmetrically distributed around the WR component (but asymmetrically around the X-ray 
component), thus causing the observed orbital modulation via electron scattering and 
photoelectric absorption.
The \ion{Si}{xiv} line absorption component is largest around phase 0.5 when the absorbing column is behind the compact star and not penetrating into the higher density wind region around the WR star. 
}

\begin{figure}
   \centering
   \includegraphics[width=9cm]{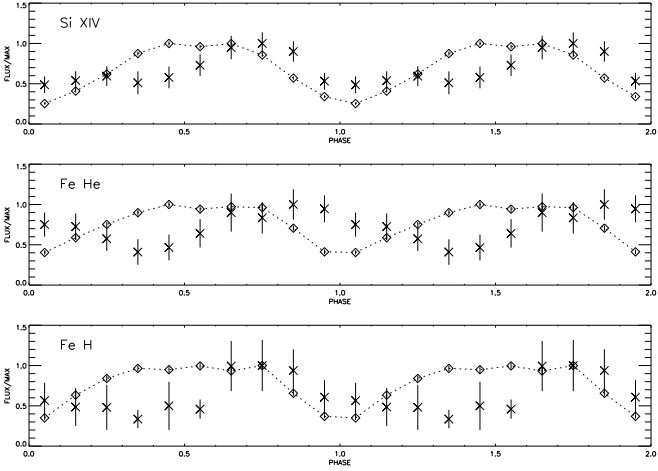}
      \caption{  { Scaled continuum fluxes (diamonds joined by  dotted lines) and  line
      fluxes (crosses, handled as single gaussians) as a function of the orbital phase for \ion{Si}{xiv} (upper), Fe He-like (middle) and Fe H-like (lower). Phases 0.0 and 0.5 correspond to the cases when the WR star and compact star are in front, respectively.}}
         \label{FeHeradvel}
   \end{figure}

  {
To study the iron lines (where no P Cygni analysis could be used) in the context of the \ion{Si}{xiv} line,
Fig.~\ref{FeHeradvel} was constructed using single gaussian fits with ten phase bins. 
As can be seen, the light curves for the lines are rather similar, suggesting the 
same absorbing agent  in all three cases. 
This might be a combination of continuum opacity for line photons, such as electron scattering, 
plus broad line absorption probably in the underlying continuum originating from the X-ray source.  
}


  \section{Constraints on the system masses and inclination}

Hanson et al. (\cite{hanson}) derive the  mass function of the compact star 
\\

 f$_C$ = 0.027 $\pm{0.01}$ M$_{\sun}$ = M$_C$$^3$$\times$sin(i)$^3$/(M$_C$+M$_{WR}$)$^2$ ,  
\\
 
\noindent assuming that the infrared \ion{He}{i} absorption line at 2.06 $\mu$m originates in the base 
of the expanding wind (K = 109 $\pm{13}$ km/s) with expected maximum blueshift around 
phase 0.75, when the WR star is approaching).
 
This mass function can be combined with the mass function we derived  from the \ion{Fe}{xxvi} line 
above:  
  \\
 
  f$_{WR}$ = 1.51$^{+1.75}_{-0.98}$ M$_{\sun}$ = M$_{WR}$$^3$$\times$sin(i)$^3$/(M$_C$+M$_{WR}$)$^2$ .  
\\
 
  For the mass ratio we obtain M$_{WR}$/M$_C$ = 3.8$^{+1.7}_{-1.4}$. 
Constraints for the compact star mass and inclination are approximately 
  \\
  
  M$_C$ = 2.8$^{+1.2}_{-1.0}$$\times$(M$_{WR}$/10) M$_{\sun}$  and 
  \\
  
  i = (38 $\pm{12}$) $\times$(M$_{WR}$/10)$^{0.33}$ degrees.
  \\
  
\begin{figure}
   \centering
   \includegraphics[width=9cm]{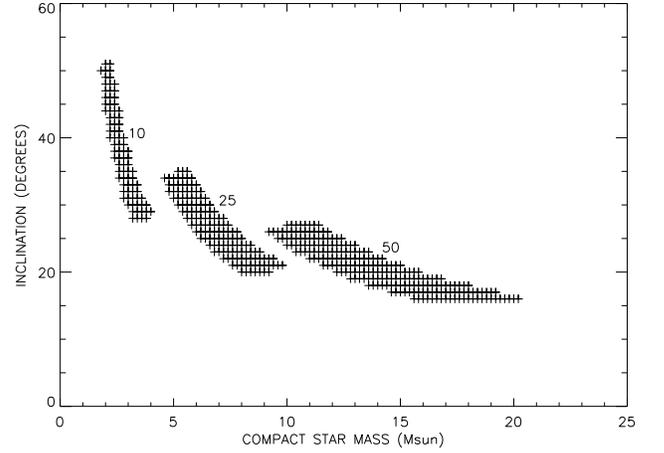}
      \caption{Permitted (compact star mass, inclination)-ranges for three WR-companion masses marked (in M$_{\sun}$) using  mass functions from the \ion{Fe}{xxvi} (this study) and \ion{He}{i} (Hanson et al. \cite{hanson}) lines. }
         \label{limits}
   \end{figure}

   More mass increases orbital
motion; hence a massive system requires smaller inclination to account for
the radial velocity amplitude. 
   As an example, if M$_{WR}$ = 60 M$_{\sun}$ then a possible solution (within the error limits) can be M$_C$ = 25 M$_{\sun}$ with a small inclination of 15 degrees.  
This is  
illustrated in Fig.~\ref{limits} where the permitted (M$_C$, inclination) ranges are shown for three  WR masses.
  
Inclinations as small as 30 degrees are acceptable
for characterizing the amplitude of the  \ion{Si}{xiv} line. 
However, the inclination can not be constrained in this way since smaller inclination can 
be compensated for by larger wind velocity (see Table~\ref{radvelcomp}). 
For large wind velocities the shape of the  continuum light curves is not very sensitive on 
inclination either. 
The effect of mass ratio on light curves is almost negligible at high wind velocities. 
  
  
A small inclination is supported by  population synthesis results on core He-burning wind-fed 
He+BH binaries by Lommen et al. (\cite{lommen}). 
They estimate that there is one such system in the Galaxy with an orbital period close 
to that of Cyg X-3 and satisfying the disc-formation criterion (11 M$_{\sun}$ + 5 M$_{\sun}$, 
compare with our Fig.~\ref{limits}). 
The existence of a disc in Cyg X-3 is supported by the  observed superluminal jets. 
Lommen et al. (\cite{lommen}) consider unlikely the existence of similar WR+NS binaries with discs, 
as already shown by Ergma and Yungelson (\cite{ergma}), due to the large wind velocity and small angular momentum of the WR+NS wind that hinder disc formation.
On the other hand, Mart\'{i} et al. (\cite{marti}) detected a bipolar jet flow at a large angle 
(73 degrees) to the line of sight. 
If the flow is parallel to the orbital plane axis, then this favours large inclination. 
However, this need not be the case for a wind-disc scenario 
as discussed by Mart\'{i} et al. (\cite{marti}) who assume an inclination of 30 degrees
(see their Fig. 7). 
If this is indeed the case, then small inclination is favoured.
  
Gladstone et al. (\cite{glad}) consider a black hole signature to be a source penetrating 
into the ultrasoft domain in their colour-colour diagram (where Cyg X-3 never entered). 
However, if the mass of the black hole is small, then its disc is hot, 
keeping it out of the ultrasoft domain populated by more massive black holes. 
On the other hand, based on spectral evolution vs. the Eddington luminosity,  
Hjalmarsdotter et al. (\cite{hjal}) point to a  massive (30 M$_{\sun}$) black hole, 
and consequently a very small inclination, with a massive WR companion. 
     
\begin{figure}
   \centering
   \includegraphics[width=9cm]{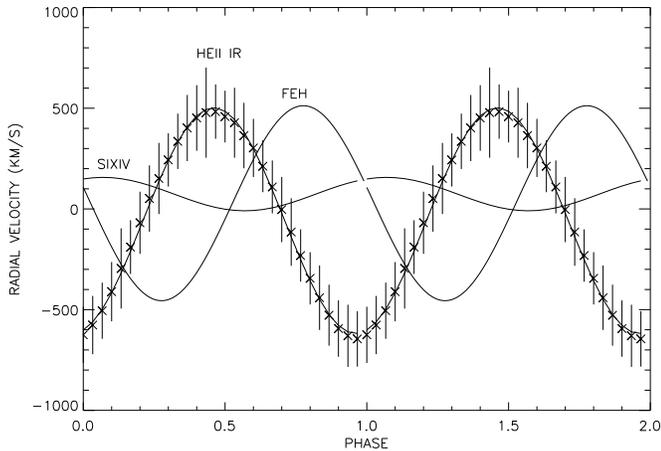}
      \caption{Model (q=5, V=2000 km/s) radial velocity curve of the infrared \ion{He}{ii} 2.09 $\mu$m line for inclination 30 degrees. The line is formed in the X-ray shadow cone behind the WR component where the wind is photoionized by the EUV radiation from the WR star. The fits for the observed \ion{Si}{xiv} and \ion{Fe}{xxvi} (H like)  lines are shown for comparison. These lines are formed in the X-ray photoionized wind and at the compact star, respectively. }
         \label{radvel3lines}
   \end{figure}

Support for low inclination in our photoionized wind model  comes from the infrared. 
We computed the emissivity and radial velocities of the IR \ion{He}{ii} 2.09 $\mu$m 
emission line (using atomic data provided by 
T. Kallman, private communication). 
It turns out that most of the line emission comes from the X-ray shadow 
(see Fig.~\ref{emissi}) where the WR star is the only ionizer. 
The radial velocity K-amplitude is around 500 km/s with maximum blueshift at phase 0.0, i.e. 
180 degrees from that of the \ion{Si}{xiv} emission,  as observed  by Hanson et al.  (\cite{hanson}). However, in our model with a wind velocity of 2000 km/s,
q=5 and inclination 60 degrees, the computed K value is much larger than that observed (1020 km/s with maximum blueshift at phase 0.0). 
If the inclination is 30 degrees, the resulting amplitude of 580 km/s would be closer to that observed 
(see Fig.~\ref{radvel3lines}) and the permitted ranges of the WR and compact star masses are 10 -- 40 M$_{\sun}$ and 2.8 -- 8 M$_{\sun}$, respectively. 
 For an inclination of 60 degrees the same ranges are 5 -- 8 M$_{\sun}$ and 1.0 -- 3.2 M$_{\sun}$.

\section{Discussion}

  {Our mass estimates (see Fig. ~\ref{limits}) corroborate
the Hanson et al. (\cite{hanson}) conclusion that if the compact component is a neutron star 
with inclination greater than 60 degrees, then the WR mass must be in the range 5 -- 11 M$_{\sun}$. 
Stark and Saia (\cite{stark}) gave an upper limit of 3.6 M$_{\sun}$ for the compact star mass using the He-like \ion{Fe}{xxv} line with a similar radial velocity curve to ours (see Fig. 7) -- however,
we prefer  to use the H-like  \ion{Fe}{xxvi} line here and 
defer a more in-depth 
discussion of the He-like  \ion{Fe}{xxv} line to a separate paper. 
They also observed  radial velocity curves of the \ion{Si}{xiv} 
and \ion{S}{xvi} lines very similar to ours. 
The difference is in their larger Gamma (recession) velocities which are due 
to ignoring the P Cygni absorption and treating 
the lines as pure Gaussians.} 

  {The formation mechanism of the \ion{Fe}{xxvi} line at 1.78 $\AA$ remains 
unknown, although it was proposed above to arise in the vicinity of the compact object.
If the  line  is formed in a hot collisionally dominated plasma with temperature T = 10 KeV, the required emission measure EM = n$^2$V is around 10$^{60}$ cm$^{-3}$ (for a distance of 9 kpc) using the MEKAL model in XSPEC. 
The emitting plasma can be e.g. a spherical corona with radius 10$^{10}$ cm (roughly the assumed accretion radius) and electron density  5$\times$10$^{14}$ cm$^{-3}$. 
The corona itself can also be a composite of many impulsive flares.  
Another alternative is that the hard coronal flux is photoionizing the disc providing 
additional Fe emission.}

In our particle simulations we assumed M$_{WR}$ = 10 M$_{\sun}$ and one may ask how a larger WR mass would influence the model results. 
A larger  mass increases the  separation and, hence, lowers the wind density in  line forming regions. Keeping all other parameters fixed, the major effect of this would be the decreasing  of  
model line luminosities in Table~\ref{lineint}. 
Unfortunately, the masses of WR stars have a large spread even for the same 
spectral type (see Fig. 4 of Crowther \cite{crowther}). 
 
   {
 The question of the origin of line absorption (Section 5.3)
is interesting, even though not fully solved here. 
The similarity of the \ion{Si}{xiv}, \ion{Fe}{xxv} and \ion{Fe}{xxvi}  light curves (Fig. ~\ref{FeHeradvel}) 
is indicative of
the same absorption agent for all. 
The  P Cygni treatment of \ion{Si}{xiv} (Fig. ~\ref{taus}) leads us to
speculate
that line photons are absorbed in the same fashion as the continuum. 
In addition, there can be  line absorption operating in the continuum (below the line photons)
that is blueshifted and varies with the orbital phase. 
A combination of these two absorptions might then cause the broad minima seen 
in  Fig. ~\ref{FeHeradvel}.  We leave a deeper discussion on line absorption for a separate paper.
 }  
  
\section{Conclusions}

 We have shown that the \ion{Si}{xiv} (6.185 \AA)  and  \ion{S}{xvi} (4.73\AA) Ly$\alpha$ emission lines of Cyg X-3, using $\it{Chandra}$  HETG observations,
 have  P Cygni-type orbitally modulated line profiles.   {The emission components  arise in the wind of the WR-companion, photoionized and excited by a hybrid source, which we take to be  composed of
 the compact X-ray source and WR He-star.  }
 
 The  radial velocity curves of the emission components have  maximum blue-shifts at phase 0.5,
 when the compact star is in front, as predicted by the wind model. 
  The K-amplitudes depend mostly on the wind velocity and constraints on the mass ratio and orbital inclination are  weak.
 The absorption component velocities are  constant around $-$900 km/s. 
    {Both the absorption and the emission components have intensity maxima (integrated gaussians) around phase 0.5,
   similar to the  continuum flux. }

  The \ion{Fe}{xxvi} Ly$\alpha$ line at 1.78 $\AA$  appears to arise in the vicinity of the compact star, as maximum blue shift occurs at phase 0.25 which is when the compact star is approaching.
   Combining the mass function derived from  this line with the mass function  from the IR \ion{He}{i} absorption line 
   (Hanson et al. \cite{hanson})  we were able to constrain the system masses and inclination.    
   Our photoionization simulations of the IR \ion{He}{ii} 2.09 $\mu$m emission line (formed in the X-ray shadow) lend 
   support to  a small orbital inclination (30 degrees) for which the range of masses of the compact 
   star is 2.8 -- 8.0 M$_{\sun}$ 
   For a larger inclination of 60 degrees this range would be 1.0 -- 3.2 M$_{\sun}$.

In the light of our results, we either have a neutron star and low-mass Wolf-Rayet star in a system of large orbital 
inclination or a black hole whose mass depends on the Wolf-Rayet mass and the system orbital inclination.

\begin{acknowledgements}
     We thank Dr Tim Kallman for advice in using the XSTAR code and providing
     additional  IR line atomic data.  DCH is grateful to the Academy of Finland for a Fellowship.   {We thank the anonymous referee for pushing us to improve the presentation.} This research has made use of data obtained from the $\it{Chandra}$  $\it{Data}$ $\it{Archive}$ and software provided by the $\it{Chandra}$ $\it{X-ray}$ $\it{Center}$ (CXC) in the application packages $\it{CIAO}$, $\it{ChIPS}$, and $\it{Sherpa}$.
\end{acknowledgements}

\end{document}